\documentstyle[amstex,aps,manuscript,floats]{revtex}

\input{psfig}
\def\Journal#1#2#3#4{{#1} {\bf #2}, #3 (#4)}

\def\NPB{{\em Nucl. Phys.} B}

\def\PRD{{\em Phys. Rev.} D}

\newcommand{\R}{{\Bbb R}}
\newcommand{\s}{{\Bbb S}}
\newcommand{\Z}{{\Bbb Z}}

\newcommand{\Kc}{{\cal K}}

\newcommand{\Rc}{{\cal R}}

\newcommand{\Tc}{{\cal T}}

\newcommand{\Kw}{{\widehat{K}}}

\begin{document}

\preprint{Imperial/TP/97-98/2}

\title{Covariant path integrals and black holes}
\author{F. Vendrell and M.E. Ortiz}
\address{Blackett Laboratory, Imperial College, Prince Consort Road, \\
London SW7 2BZ, U.K.}

\maketitle

\begin{abstract} 
The thermal nature of the propagator in a collapsed
black-hole spacetime is shown to follow from the non-trivial
topology of the configuration space in tortoise coordinates by using
the path integral formalism.
\end{abstract}
\mbox{} \hspace{1mm}

The path integral (PI) formalism is useful to calculate propagators in 
configuration spaces (CS) endowed with a non-trivial topology, such as
in curved spacetimes.  Moreover, even if the CS topology is trivial, 
this may not be the case for its Euclidian section, where the PI
should always be computed.  For example, in an eternal black-hole background 
or in Rindler spacetime, although the CS itself has a trivial topology in 
Kruskal or Rindler coordinates, the Euclidian CS has a periodic structure in 
the Euclidian time-like coordinate.  In these spacetimes endowed with an 
event-horizon (EH), it can be shown that the thermal properties of the 
propagator follows from the periodic structure of the Euclidian 
CS \cite{HH,tro,gui,fl}.

In a collapsing black-hole spacetime, however, this periodic structure is
missing.  One has to find in this case another procedure to obtain the 
thermal properties of the propagator. It is the purpose of this paper to show 
that a similar periodic structure may be recovered in tortoise coordinates if 
one requires that they cover the entire spacetime by allowing them to take 
complex values. In the resulting complex CS, the EH has a cylinder-like 
topology. The propagator is then obtained from a PI by adding all the 
contributions of the classes of path whith different winding number around 
the EH \cite{OV}. An advantage of this procedure is that the contributions of 
the paths crossing the EH are also included in the PI, which is not the case 
if the Euclidianisation is performed in Kruskal coordinates.

One of the simplest models for gravitational collapse is the Synge model, 
where a Schwarz\-schild black hole is created from an imploding spherical 
shell of radiation~\cite{Sy}.  
We shall concentrate our attention on the spacetime region $\Rc$
where the thermal radiation emitted by the black hole may be 
detected by an inertial observer.  This region is defined to be located 
outside the imploding shell, far from the black hole and at late times.
If $x$ denotes the Kruskal coordinates, this region is defined by $x^+\gg1$ and
$x^-\approx x^-_H$, where $x^-_H\equiv x^+_0-4M$ is the position of the EH.
In this region, the line element is given by
\begin{eqnarray}
ds^2 \approx \frac{dx^+\,dx^-}{\kappa\left(x^-_H-x^-\,\right)} 
- r(x^+,x^-)^2\,\left(\,d\theta^2+\sin^2\theta\,d\phi^2\,\right),
\label{ds0x}
\end{eqnarray}
where $\kappa=(4M)^{-1}$, and the tortoise coordinates $y$ are defined by
\begin{eqnarray}
\begin{array}{rcccccl}
x^+(y^+) &\approx& y^+, &\qquad&
x^-(y^-) &\approx& x^-_H - 2e^{-\kappa (y^--x^-_H)},
\label{x0(y)}
\end{array}
\end{eqnarray}
where $(y^0,y^1)= (t,r+2M\ln\vert r-2M\vert)$, if $t$ and $r$ are the 
Schwarzschild coordinates.  The Kruskal CS is given by
\begin{eqnarray}
\Kc=\{\,(x^+,x^-,\theta,\phi)\in\R^2\times\s^2 
\ \vert\ r(x^+,x^-)\geq 0\,\},
\end{eqnarray}
and its homotopic properties are trivial.

\begin{figure}
\psfig{figure=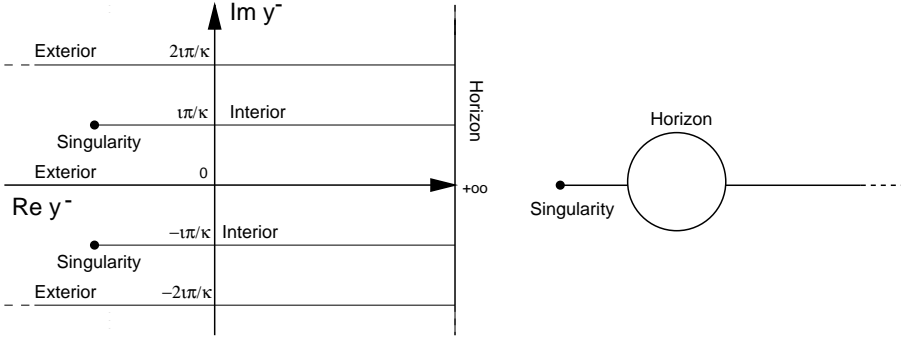,height=1.8in}
\mbox{} \vspace{1mm}
\caption{A section of the tortoise complex covering space (on the left) and
of the configuration space $\Tc$ (on the right) along the $y^-$ coordinate.
\label{fig}}
\end{figure}

The tortoise coordinates cover only the region $x^-<x^-_H$ outside the EH,
and so cannot be used to parameterise a path crossing the EH.
Instead of introducing another set of coordinates to cover this region,
one extends tortoise coordinates to {\it complex} values, to keep a sense 
of connectedness for the CS.  This is possible since
$x^-(y^-+i\pi/\kappa)>x^-_H$, and thus the complex number $y^-+i\pi/\kappa$ 
describes a point inside the EH.  The CS $\Tc$ in tortoise coordinates is 
thus defined as the complex preimage of the CS $\Kc$ through the 
transformation $x=x(y)$.  Since the transformation $x^-=x^-(y^-)$ is periodic 
in the imaginary direction, a point $x\in\Kc$ has a infinite number of complex 
preimages denoted by $y^\nu$, where $\nu\in\Z$. They are given by 
$\left(y^\nu\right)^+=x^+$ and
\begin{eqnarray}
\left(y^\nu\right)^-= \left\{
\begin{array}{lr}
y^- + i2\pi(\nu+1/2)/\kappa, \quad & \mbox{{\it if} \ $x^->x^-_H$,}  \\ [2mm]
y^- + i2\pi\nu/\kappa, & \mbox{{\it if} \ $x^-<x^-_H$,} 
\end{array} \right.
\label{ynu}
\end{eqnarray}
where $y^-$ is the real preimage of $x^-_H-\vert x^-_H-x^-\vert$.
The complex CS $\Tc$ is thus given~by
\begin{eqnarray}
\Tc = \{\,(y^+,y^-,\theta,\phi)\in 
\R\times \left[\,\bigcup_{\mu\,\in\,\Z} R_{\mu/2}\bigcup \, T\,\right]
\times\s^2 \ \vert\ r(y^+,y^-)\geq0 \,\},
\end{eqnarray}
where $R_\mu=\R+i2\pi\mu/\kappa$ and $T = +\infty + i \R$ (the set $T$ 
parameterises the EH).  The CS $\Tc$ is multiply connected and is represented 
in fig.~\ref{fig}.  The topology of the EH in $\Tc$ is given by
$\R\times\s^1_{1/\kappa}\times\s^2$, where $\s^1_{1/\kappa}$ is a
circle of radius $\kappa^{-1}$.  The fundamental group of $\Kc$ is
thus isomorphic to $\Z$.  The classes of paths are labelled by the
integer $\nu$ which is the winding number of the paths around the
circle $\s^1_{1/\kappa}$.  

In quantum mechanics, the amplitude to move from an initial point
$x_i$ to a final point $x_f$ in a parameter-time
$s=s_f-s_i$ is given by the propagator $K(x_i,x_f;s)$, or heat kernel.
It is defined as a PI with endpoints $x_i$ and $x_f$, and is 
written symbolically as
\begin{eqnarray}
K_{{\mathrm Kruskal}}(x_i,x_f;s) =
\sideset{}{_g}\sum_{x(\cdot)\,\in\,\Kc}
e^{\frac{i}{\hbar}S_g[\,x(\cdot)\,]},
\label{SPX}
\end{eqnarray}
where $\sideset{}{_g}\sum$ and $S_g$ are the covariant sum over paths and 
action respectively \cite{BP}.  The set of paths on which the sum is taken is 
fixed by the CS $\Kc$ and defines the {\it Kruskal vacuum}.   
If both endpoints $x_i$ and $x_f$ belong to the region $\Rc$, one may try
replacing the metric $g$ appearing in the path integral {\it everywhere} 
by the metric of Eq.~(\ref{ds0x}).  The weight of each path that ventures out 
of the region $\Rc$ is changed, but we may conjecture that this will not 
greatly affect the propagator in $\Rc$, since this is a local quantity.
The simplicity of choice of the metric (\ref{ds0x}), which is the Minkowski
metric $\eta$ in tortoise coordinates, allows us to compute the propagator 
easily.  Because the propagator is a biscalar, one writes
\begin{eqnarray}
\sideset{}{_g}\sum_{x(\cdot)\,\in\,\Kc} \ 
e^{\frac{i}{\hbar} S_g[\,x(\cdot)\,]} 
\approx \sideset{}{_\eta}\sum_{y(\cdot)\,\in\,\Tc} \ 
e^{\frac{i}{\hbar} S_\eta[\,y(\cdot)\,]},
\label{SPxy}
\end{eqnarray}
and thus the sum may be taken over paths in the complex tortoise CS.
The PI may be rewritten by summing over the classes of paths.  The 
contribution of the entire class of paths with winding number $\nu$ gives the 
free propagator $\Kw_0$ (in tortoise coordinates) with arguments $y_i$ and 
$y_f^\nu$. The total propagator between $y_i$ and $y_f$ (both in region $\Rc$)
is given by
\begin{eqnarray}
\Kw_{{\mathrm Kruskal}}\left(\,y_i; y_f;\,s\,\right)
\approx \sum_{\nu\in\Z} \Kw_0
\left(\,y^+_i,y^-_i,\Omega_i \,; 
y^+_f, y^-_f+i2\pi\nu/\kappa,\Omega_f\,; s\,\right),
\label{final}
\end{eqnarray}
and represents an outgoing thermal flux of particles with temperature 
$T=\hbar\kappa/(2\pi k)$.  

We have shown that the homotopic properties of the
configuration space are not an intrinsic feature of the black-hole
spacetime, but depend on the set of coordinates chosen to cover it.
Since the inverse of the transformation relating Kruskal and tortoise
coordinates is not analytic, the CS $\Kc$ and $\Tc$ have different
topologies. These properties have no physical consequences on the classical 
motion of a particle. However, in a quantum mechanical framework, a particle
may tunnel across the horizon, and the topology of the whole
configuration space is then physically relevant.  The radiation of a
black hole is thermal because, from the point of view of a distant
inertial observer, there is a denumerably infinite number of ways for
a particle to tunnel through the horizon.  

\section*{Acknowledgments}

F.V.~acknowledge support from the Swiss National Science Foundation.


\end{document}